\documentclass[iop,superscriptaddress]{emulateapj}
\usepackage{amsmath,epsfig,url,natbib}
\usepackage{graphicx,epstopdf,float,color,array}
\usepackage{graphicx}
\usepackage{multirow}
\usepackage{rotating,multirow}
\usepackage{hyperref}
\usepackage{wrapfig}
\usepackage{float}
\usepackage{appendix}
\hypersetup{colorlinks=true, citecolor=blue, filecolor=magenta,
  urlcolor=cyan, linkcolor=blue}

\definecolor{orange}{cmyk}{0,0.5,1,0}

\definecolor{orange}{cmyk}{0,0.5,1,0}

\shorttitle{Resolved properties: CANDELS+MUSE }
\shortauthors{Jafariyazani et al.}

\begin{document}

\title{Spatially resolved properties of galaxies from CANDELS+MUSE: Radial extinction profile and insights on quenching}

\author{Marziye Jafariyazani\altaffilmark{1},
Bahram Mobasher\altaffilmark{1}, Shoubaneh Hemmati\altaffilmark{2}, Tara Fetherolf\altaffilmark{1}, Ali Ahmad Khostovan \altaffilmark{3}, Nima Chartab\altaffilmark{1}}

\email{mjafa003@ucr.edu}
\altaffiltext{1}{Department of Physics and Astronomy, University of California, Riverside, CA, USA}
\altaffiltext{2}{Jet Propulsion Laboratory, California Institute of Technology, Pasadena, CA, USA}
\altaffiltext{3}{NASA/Goddard Space Flight Center, Astrophysics Science Division, Greenbelt, MD, USA}

\begin{abstract}

Studying the internal processes of individual galaxies at kilo-parsec scales is crucial in enhancing our understanding of galaxy formation and evolution processes. In this work, we investigate the distribution of star formation rate (SFR), specific SFR (sSFR), and dust attenuation across individual galaxies for a sample of 32 galaxies selected from the MUSE-Wide Survey at 0.1 $< \textit{z} <$ 0.42 with a dynamic range in stellar masses between $10^{7.7}$ and $10^{10.3}$ M$_{\odot}$. We take advantage of the high spatial resolution of the MUSE integral-field spectrograph and measure reliable spatially resolved H$\alpha$ and H$\beta$ emission line maps for individual galaxies. We also derive resolved stellar mass, SFR and dust maps using pixel-by-pixel SED fitting on high resolution multi-band \textit{HST}/ACS and \textit{HST}/WFC3 data from the CANDELS survey. By combining these, we analyze the radial profile of various physical parameters across these galaxies. We observe a radial dependence in both stellar and nebular color excess profiles peaking at the inner regions of galaxies. We also find the color excess profiles to most strongly correlate with the integrated sSFRs of galaxies. The median sSFR$_{\mathrm{H}\alpha}$ radial profiles of galaxies in our sample show a 0.8 dex increase from the central regions outward. This increase compared to the almost flat median radial profile of sSFR$_{\mathrm{SED}}$, which traces longer timescales of star formation, is in favor of the inside-out quenching of star formation. We bring further evidence for this quenching scenario from the locus of different subregions of galaxies on the SFR-M$_{*}$ and sSFR-M$_{*}$ relations.

\end{abstract}


\section{Introduction\label{sec:intro}}

Over the past decades, great deal of work has been done to understand the physical properties of galaxies over a wide range of cosmic time. However, most of these studies are based on measurements of the integrated light of galaxies and most information about the behavior of galaxies at small scales (kpc level) are ignored. Resolved studies of local galaxies, including the Milky Way, show complex distributions of metalicities, stars, gas, dust, and star formation activity (e.g., \citealp{sanchez2012,Dalcanton2012,Bundy2015}). This suggests that to draw a comprehensive picture of how galaxies evolve, what processes are involved, and how such processes occur requires that we study galaxies at high spatial resolution scales to understand the internal mechanisms that drive the underlying galaxy's evolution.

A relatively inexpensive and accurate method, widely used to measure physical parameters for large samples of galaxies, is to fit their Spectral Energy Distribution (SED) using multi-waveband photometric data. This method is recently extended to measure these properties at the resolution element of galaxies, providing pixel-by-pixel estimates and therefore, high resolution maps (e.g., \citealp{Wuyts2012,Hemmati2014,Guo2018}). However, relying solely on SED-derived parameters is problematic due to potential degeneracies and uncertainties associated with the SED-fitting process. Stellar mass is known to be the most reliable output from SED-fitting, since it is measured by the normalization of the SED, while properties such as star formation rates (SFR), color excess, age, etc. rely on a wide range of assumptions. For instance, SED-derived SFR is highly degenerate with age, dust, and metallicity and is dependent on the assumed star formation history (SFH) and initial mass function (IMF) (\citealt{Conroy2013}). Furthermore, physical properties should be derived from independent methods rather than deriving all from SED-fitting to properly study their correlations with each other without being worried about model-dependent correlations. Hence, using spectroscopic information besides photometry seems to be required to break degeneracies and obtain more robust results.

Another challenge in measuring physical properties of galaxies, both at integrated and resolved scales, is estimating the effect of dust on the measured properties. It is known that SED color excess is seriously affected by age - dust - metallicity degeneracy, and it only measures reddening toward the stellar light which can be different from reddening toward nebular emission lines (\citealt{calzetti1997}). This different level of attenuation toward nebular and stellar regions is explained to some extent by two-component dust model proposed by \citet{charlot2000}. In this model, old stars which are distributed all over the galaxies become redder due to the attenuation only from the diffuse interstellar medium (ISM), while young, short-lived massive stars formed in cold molecular clouds tend to ionize their surrounding regions, and are subject to attenuation from ionized gas as well. Therefore, it is important to measure both forms of reddening not only to correct their effect on other physical parameters, but also to better understand the geometry of stars and dust inside galaxies. This also requires spectroscopy to directly measure diagnostic lines  sensitive to extinction (i.e. Balmer decrement: H$\alpha$/H$\beta$) to estimate the attenuation toward nebular regions.

In recent years, spatially-resolved spectroscopy of local galaxies became feasible thanks to the integral field spectroscopy (IFS) surveys, such as  SAMI (\citealt{croom2012,scott2018}), CALIFA (\citealt{sanchez2012}), and MaNGA (\citealt{Bundy2015}). For this work, we needed IFS data in a field with available high spatial resolution multi-waveband photometry, so we take advantage of MUSE-Wide Survey (\citealt{MUSE2}) which also allowed us to go further than previous studies by examining the resolved properties of galaxies at lower masses and higher redshifts. We focus on studying radial profiles of SFR, sSFR and dust attenuation in a sample of 32 galaxies with high S/N detection of H$\alpha$ and H$\beta$ at 0.1 $< \textit{z} <$ 0.42 and average stellar mass of $10^{8.71}$ M$_{\odot}$. The aim is to evaluate the effect of dust on measured properties, study the radial gradient in nebular and stellar reddening, and constrain quenching mechanisms based on radial profile of the SFR and sSFR for this sample.

The paper is organized as follows. In \S 2 we present the sample and its selection criteria. \S 3 provides our methodology in analyzing the photometric and spectroscopic data from the Cosmic Assembly Near-infrared Deep Extragalactic Legacy Survey (CANDELS) and MUSE-Wide Survey respectively at kpc scale. We discuss our results from our resolved analysis of SFR, sSFR and dust distribution in galaxies in \S 4. In \S 5 we summarize the main points of this work and discusses future directions.

Throughout this paper, we adopt a cosmology with a matter density parameter $\Omega_{M}= 0.3$, a cosmological constant $\Omega_{\Lambda}= 0.7$ and a Hubble constant of H$_{0}$ = 70 kms$^{-1}$Mpc$^{-1}$. All magnitudes are in the AB system.

\section{Sample selection}

 Accurate measurement of the physical properties of galaxies at kpc-scales requires a sample with high-spatial resolution multi-waveband photometry and integral field spectroscopy. With extensive dataset currently available for galaxies in the GOODS-South field, this provides an ideal sample as it combines the deepest and highest resolution multi-band \textit{HST} data from the CANDELS survey with integral field spectroscopic data of high spectral and spatial resolution from the MUSE-Wide Survey. In this section we briefly explain the two parent samples followed by our selection strategy.

\subsection{CANDELS}

CANDELS (\citealp{Grogin2011,koekemoer2011}) is the largest single project carried out by the Hubble Space Telescope (\textit{HST}), with 902 orbits of observing time. CANDELS consists of five fields (GOODS-South, GOODS-North, EGS, UDS, COSMOS) with the optical imaging data from the Advanced Camera for Surveys (ACS) and infrared data from Wide Field Camera 3 (WFC3). The catalog is selected in WFC3 H-band (F160W) and contains 18000 galaxies to a 5$\sigma$ limiting magnitude of $\sim$ 27. \textit{HST} images, multi-wavelength photometric catalogs and catalogs of integrated physical properties for CANDELS galaxies are now available in all five fields (for details see, \citealt{Guo2013}, \citealt{Galametz2013,Mobasher2015,Nayyeri2017,Stefanon2017}).

\subsection{MUSE-Wide Survey}
MUSE-Wide Survey is a blind 3D spectroscopic survey of sub-areas in the CANDELS-DEEP and CANDELS-COSMOS regions with 1-hour exposure time per 1 arcmin$^2$ pointing. It has been done using the the MUSE instrument on the Very Large Telescope (VLT) in the wide field mode, which provides medium resolution spectroscopy at a spatial sampling of $0.2^{''}$ per spatial pixel. The extended wavelength range was used, covering 4750-9350 \AA, with $\sim$ 2.5 \AA\, resolution. The final survey will cover 100 fields each covering an area of 1 arcmin$^2$. This paper uses data from 24 MUSE fields in the GOODS-S deep area (\citealt{herenz2017}). The data from these fields provided the preproccessed datacubes for a total of 831 emission line galaxies with redshifts in the range 0.04 $< \textit{z} <$ 6.

\subsection{Final sample}
The main aim of this paper is to measure the kpc scale distribution of SFR, sSFR, and dust in galaxies and study them as a function of their integrated physical properties. Recent star formation activity is measured using H$\alpha$ emission line, and dust distribution is estimated from Balmer decrement (H$\alpha$/H$\beta$) which provides a measure of the extinction correction. Therefore, we select the sample such that H$\alpha$ and H$\beta$ emission lines both fall in the observed wavelength range covered by the spectrograph to enable measurements of both lines simultaneously. This restricts the redshift range of our sample to $0.1 < \textit{z} < 0.42$. Finally, we omit galaxies for which their H$\alpha$ and/or H$\beta$ lines could not be fitted properly due to low S/N or being at the edge of the field of view, such that the whole galaxy was not covered. Only 32 galaxies (out of a sample of 831) satisfy our selection criteria with the stellar mass spanning the range $10^{7.7}$ to $10^{10.3}$ M$_{\odot}$ and average mass of $10^{8.71}$ M$_{\odot}$. 

\begin{figure}
\includegraphics[width=0.47\textwidth]{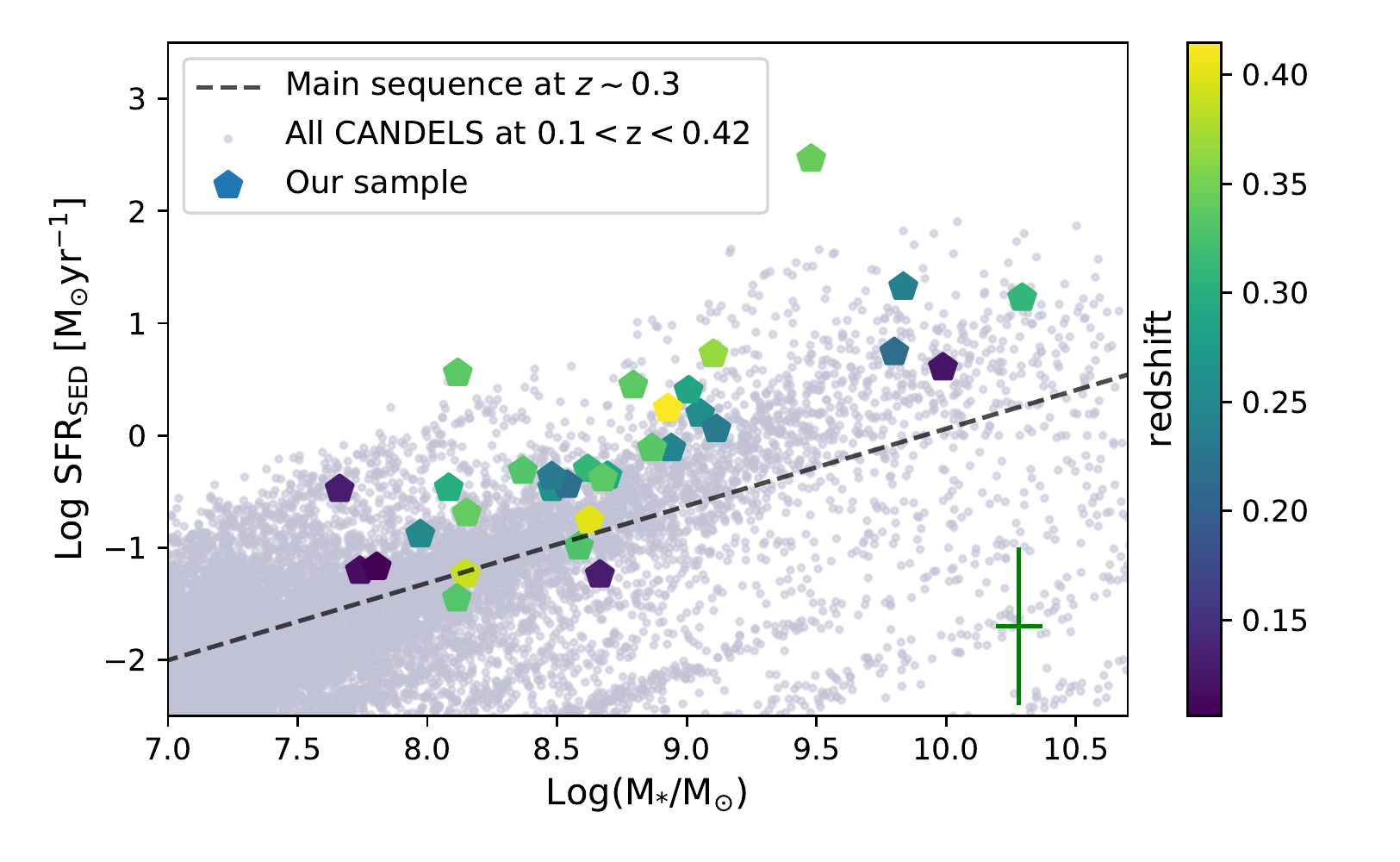}
\caption{ Our sample of 32 galaxies with MUSE detection of  H$\alpha$ and H$\beta$ lines over plotted on the galaxies from the CANDELS survey at the same redshift range: $0.1< z< 0.42$. Black dashed line is representing the main sequence of star-forming galaxies from \citealt{lee2015} for galaxies at $z \sim 0.35$. Galaxies of this sample are color-coded according to their redshift.}
\label{fig:samplems}
\end{figure}

Figure \ref{fig:samplems} shows our final sample on the SFR-M$_*$ relation, compared with all the galaxies observed in the CANDELS GOODS-S field at the same redshift range ($0.1 < \textit{z} < 0.42$). This shows how the SFR and stellar mass of galaxies in our sample compare to the overall population of galaxies at this redshift range. Stellar mass and SFR in this plot are derived from SED-fitting with details explained in section 3.1. The dashed black line represents the main sequence of star-forming galaxies at $z\sim 0.35$ from \citealt{lee2015}. Our galaxies are typically towards the upper part of the “main sequence”, suggesting we are looking at the more active population of star-forming galaxies which is expected on H$\alpha$ selected samples (e.g. \citealt{Nelson2016_2,emami2018}). Despite our small sample size, the galaxies in our sample cover $\sim 3$ orders of magnitude in stellar mass. 

\section{Measurement of resolved properties}

\subsection{Resolved photometric measurements}

In this section we describe our technique in measuring resolved photometric properties of galaxies. Following \citealt{Hemmati2014}, we perform pixel-by-pixel SED fitting using \textit{HST} imaging data in seven pass bands. The \textit{HST} imaging data consist of four optical bands from the ACS (F435W, F606W, F775W, and F850LP) and three near-infrared bands from WFC3 (F105W, F125W, and F160W) from the CANDELS GOODS-S survey. An automated pipeline is developed to perform high spatial resolution measurements. First we make a 100$\times$100 pixel cutout around each source and produce a segmentation map of that galaxy. The goal of the segmentation map, is to simply define edges for the galaxy with 0 and 1 values for pixels not corresponding and corresponding to the galaxy respectively. Multiplying the segmentation map to each of the cutouts removes the surrounding objects and background pixels. We then perform PSF matching to degrade HST images to the MUSE
resolution ($\sim$ 1.1 arcsec) at H$\alpha$ wavelength. A multi-waveband catalog is then generated for each galaxy with each row corresponding to a resolution element with magnitudes in seven pass bands and their associated RMS errors. The spectroscopic redshift for each galaxy is also listed.

Model SEDs are generated covering a wide range in parameter space. The model template library includes SEDs for stars from the PICKLES library (\citealt{Pickles}), AGN and normal galaxies. For quasars, we use synthetic and composite quasar libraries available in the LePhare package (\citealt{Arnouts1999,ilbert2006}), and for galaxies we build a unique and inclusive library using \citealt{Bruzual2003} stellar population synthesis models. We assume a \citealt{chabrier} initial mass function, exponentially declining star formation histories with $\tau$ in the range 0.05 to 20 Gyr, and the metallicity to be 40\% Solar.

SED-fitting on the resolved elements is then performed using the LePhare SED-fitting code, which finds the closest match from the model templates to the observed SED corresponding to each resolution element based on $\chi ^2$ minimization. The redshifts are fixed to their spectroscopic values to fit for the physical parameters including stellar mass, SFR, and extinction. Lephare code enables us to add the contribution of emission lines including Ly$\alpha$, H$\alpha$, H$\beta$, [O{\sc ii}], [O{\sc iii}]4959, and [O{\sc iii}]5007 based on Kennicut relations (\citealt{Kennicutt:1998zb} ) and incorporate Calzetti dust attenuation law (\citealt{CALZETTI2000}). Figure \ref{fig:resolvedmaps} (in the appendix) shows resolved stellar mass and SFR surface density as well as stellar E(B$-$V) maps for three sample galaxies. For visualization purposes, these maps show results of resolved SED-fitting on \textit{HST} quality data where images from seven photometric bands were PSF matched to the resolution in F160W. Overall the observed trends are not affected by degrading the \textit{HST} images to MUSE resolution.

\vspace{2mm}
\subsection{Resolved spectroscopic measurements}

Spatially resolved spectroscopic measurements are performed using data products from the MUSE-Wide Survey. We use 3D reduced datacubes for all galaxies in our sample with their detailed reduction procedure presented in \citet{herenz2017}. To produce emission line maps, we first fit the stellar continuum with a low-order polynomial and subtract it from each spaxel. Then the emission line fluxes are measured by fitting the spectra by a Gaussian, weighted by the variances of the spectrum for every spaxel. In measuring the integrated emission line fluxes, we do not account for any possible effect by kinematics due to both the relatively low stellar masses and the exclusion of inclined galaxies (axial ratio $ b/a < 0.5 $) from the sample.

Using our emission line maps, we then measure H$\alpha$/H$\beta$ to estimate the effect of dust. From quantum physics, we know that the flux ratio of two nebular Balmer emission lines is constant for a fixed electron temperature, and deviation from the theoretical value can be due to the dust extinction. We measure Balmer decrement pixel-by-pixel for the spaxels where both H$\alpha$ and H$\beta$ have S/N $>$ 3 to produce a reliable map for each galaxy. In this process, pixels which have an observed value of F(H$\alpha$)/F(H$\beta$) $<$ 2.86, which is the theoretical threshold for Case B recombination at T = 10000 K in the absence of dust (\citealt{Osterbrock1989}), are assigned zero extinction.

SFR$_{\mathrm{H}\alpha}$ is computed using calibration from \citet{kennicutt1998} for individual pixels, and to be consistent with our SFR$_{\mathrm{SED}}$ measurements, we adjusted them to account for our SED-fitting assumptions compared to \citet{kennicutt1998} assumptions: \citet{chabrier} IMF instead of \citet{salpeter} IMF, and 40\% Solar metallicity instead of Solar. Balmer color excess, E(H$\beta$-H$\alpha$), is computed following Equation \ref{eq:color_excess} using Balmer dercrement maps. Then A(H$\alpha$), attenuation toward HII regions,  is measured for each pixel following Equation \ref{eq:AHa}. In this Equation, k(H$\alpha)$ and k($H\beta)$ are the amount of extinction at the H$\alpha$ and H$\beta$ wavelengths, assuming a \citealt{CALZETTI2000} dust attenuation law. Finally, SFRs per pixel are corrected for extinction using Balmer decrement.

\begin{equation}
E(H\beta-H\alpha)= 2.5\;\log_{10} \frac{(H\alpha/H\beta)_{obs}}{2.86}
\label{eq:color_excess}
\end{equation}

\begin{equation}
A(H\alpha)=\frac{E(H\beta-H\alpha)}{k(H\beta)-k(H\alpha)}\times k(H\alpha)
\label{eq:AHa}
\end{equation}

\begin{figure}
\includegraphics[width=0.47\textwidth]{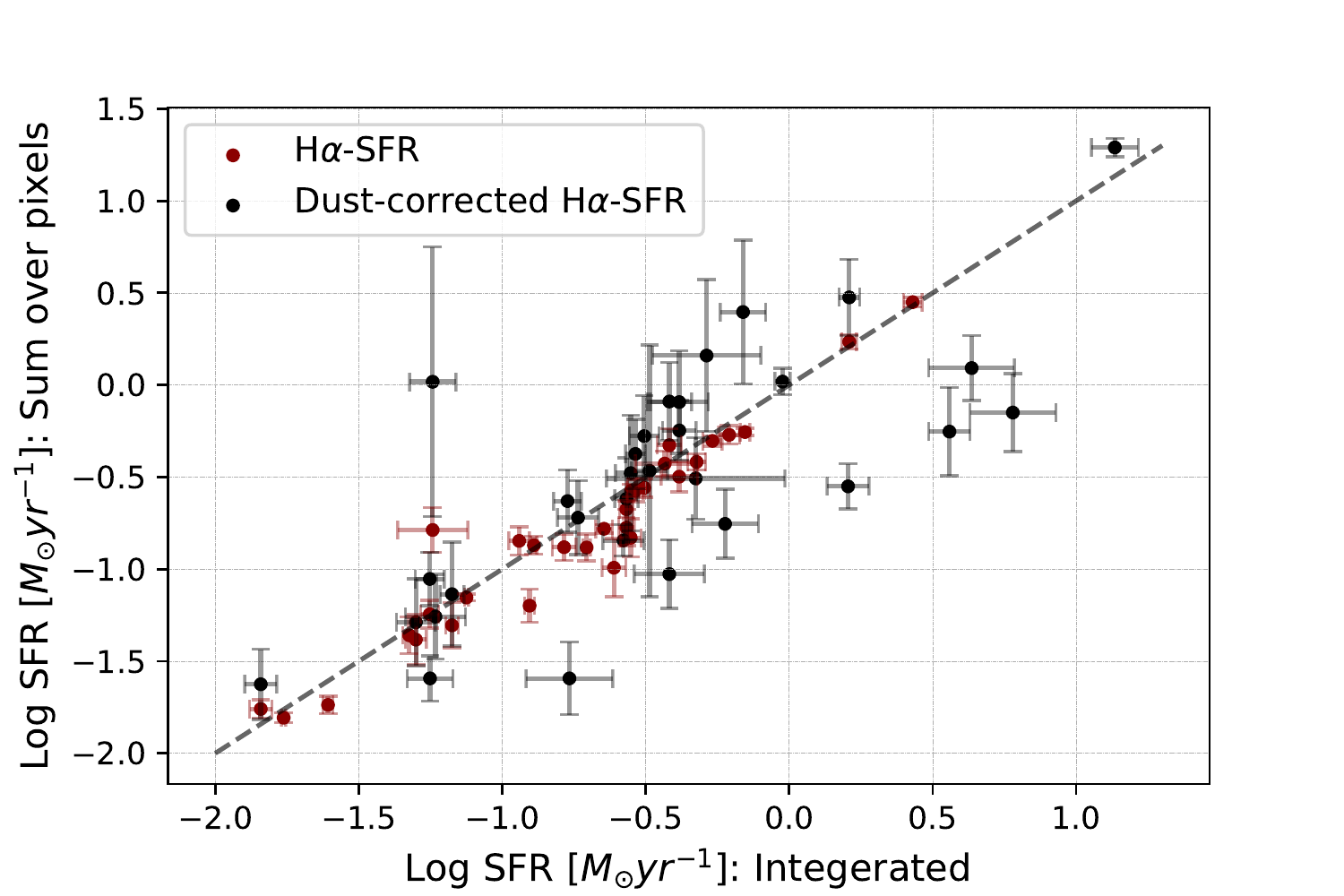}
\caption{Total SFR of galaxies measured by summing up the SFRs per pixel vs. SFR measured from the integrated emission lines. Red data points are SFR$_{\mathrm{H}\alpha}$ and black ones are dust-corrected SFR$_{\mathrm{H}\alpha}$. Dashed line represents one-to-one relationship.}
\label{fig:sfrcompare}
\end{figure}

As a sanity check, we examined if SFR computed from the integrated emission line of the galaxies are equivalent to the sum of the SFRs of their pixels, as shown in Figure \ref{fig:sfrcompare} for the case of observed and dust-corrected SFR$_{\mathrm{H}\alpha}$. We find that the measurements are in good agreement such that both independent approaches are consistent. The only concern is that incorporating lower signal to noise H$\beta$ measurements significantly increased the error bars and scatter between the two SFRs.

\newpage

\section{Results}

\subsection{Integrated SFR diagnostics}

\begin{figure}[!htb]
\includegraphics[width=0.47\textwidth]{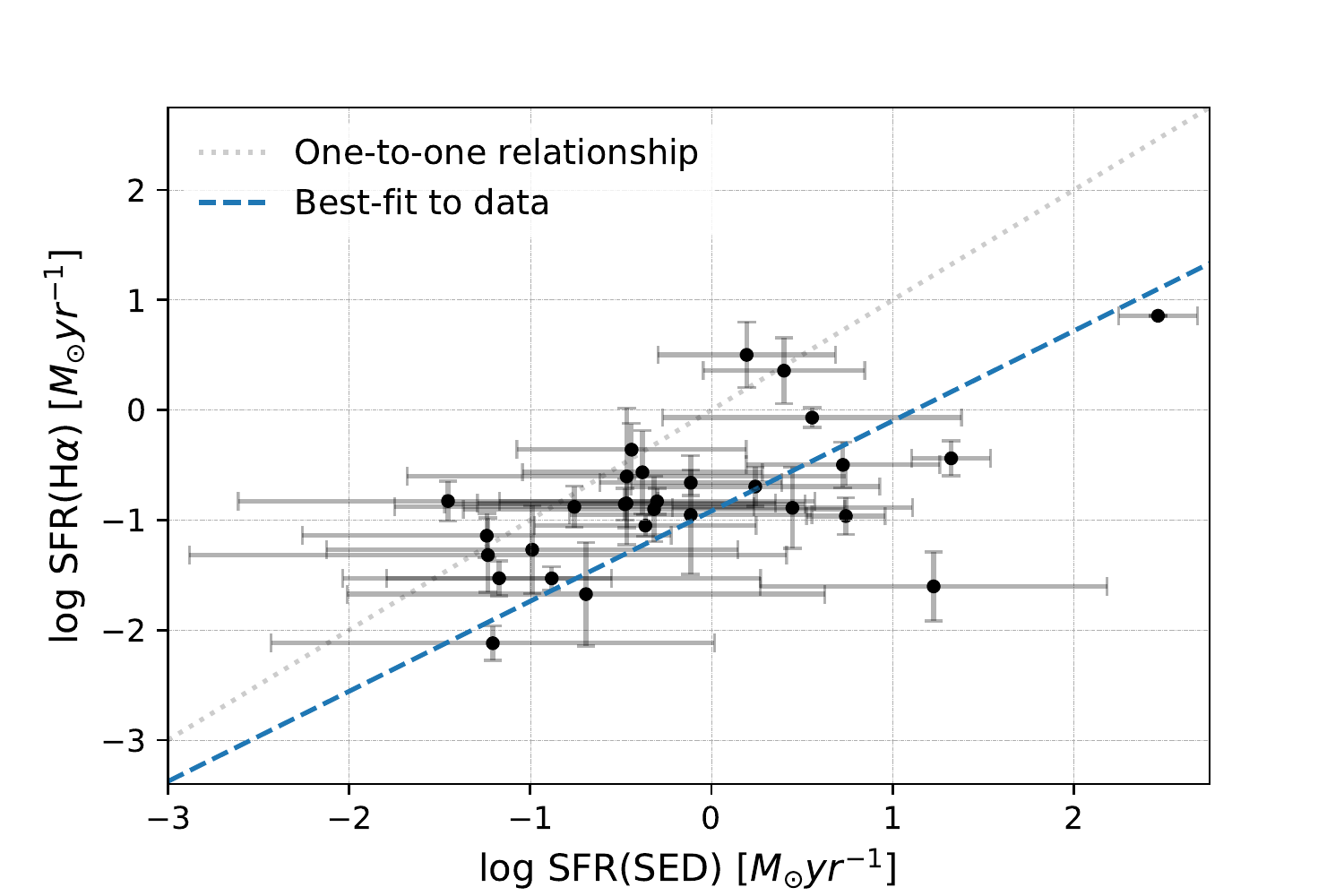}
\caption{Comparison between the SFR$_{\mathrm{H}\alpha}$ and SFR$_{\mathrm{SED}}$ for the integrated light of galaxies. Gray dotted line represents the one-to-one relationship. Blue dashed line represents the best fit to the data with the slope $=0.819\pm 0.232$, and intercept$=-0.919\pm 0.152$.}
\label{fig: TwoSFRcompare}
\end{figure}

We start by comparing the dust-corrected $\rm SFR_{H\alpha}$ and $\rm SFR_{SED}$ for the integrated light of galaxies in our sample. $\rm SFR_{H\alpha}$ is tracing the ionizing radiation from massive stars ($>10M_{\odot}$) which provides a relatively instantaneous measure of the SFR ($<10 Myr$) (\citealt{Kennicutt:1998zb}) whereas SFR$_{\mathrm{SED}}$ is measuring the average SFR of the galaxy over its lifetime based on the assumed SFH. Despite tracing different star formation timescales and large errors in SFR$_{\mathrm{SED}}$, these two diagnostics are well-correlated (p-value= 0.0004) as shown in Figure \ref{fig: TwoSFRcompare}. The blue dashed line in Figure \ref{fig: TwoSFRcompare} shows the best-fit linear relationship, where the slope is $0.819\pm 0.232$ and the intercept is $-0.919\pm 0.152$. The slope of this relation is close to unity for lower SFRs with deviation from unity at higher SFR values. This is consistent with the results from \citet{shivaei2016} who found a near unity relation between dust-corrected SFR$_{\mathrm{H}\alpha}$ and SFR$_{\mathrm{SED}}$ (UV-to-FIR) for a sample of z $\sim$ 2 galaxies. We note that, while the SFR$_{\mathrm{H}\alpha}$ is reliably corrected for dust attenuation by the Balmer decrement, not including the FIR data in the SED fitting might produce slight underestimation of SFR$_{\mathrm{SED}}$ at larger SFR values where more dust attenuation is expected (e.g., \citealt{Reddy2010,Casey2014,shivaei2016}). This increases the deviation of the two measures at higher SFRs (decreasing the slope). 

Our sample of emission line galaxies are residing on or above the main sequence of star forming galaxies (e.g., \citealp{Noeske2007,Elbaz2011}), with more massive objects having higher average SFR values (see Figure \ref{fig:samplems}). The deviation of the two SFR diagnostics (probing different SFR timescales) at larger SFRs, is hence suggestive of the quenching process starting in the more massive objects in our sample (the so called mass quenching; \citealp{Peng2010}). In the following subsections we look into resolved distribution of SFR, sSFR, and dust to further learn about the quenching mechanism. 

\subsection{Radial profiles of mass, SFR and sSFR}\label{sec:sfr profile}

To combine and compare the resolved spectroscopic and photometric products we transform our pixel measurements to radial profiles by taking the median value of all the pixels in each annulus. We then use the bootstrap method to estimate the standard error of the sample median. While this technique loses some information (e.g. smooths out clumps present in galaxies), the strategy facilitates the interpretation of the results as we avoid complexity and uncertainties of matching pixels between two sets of data.

\begin{figure}
\includegraphics[width=0.47\textwidth]{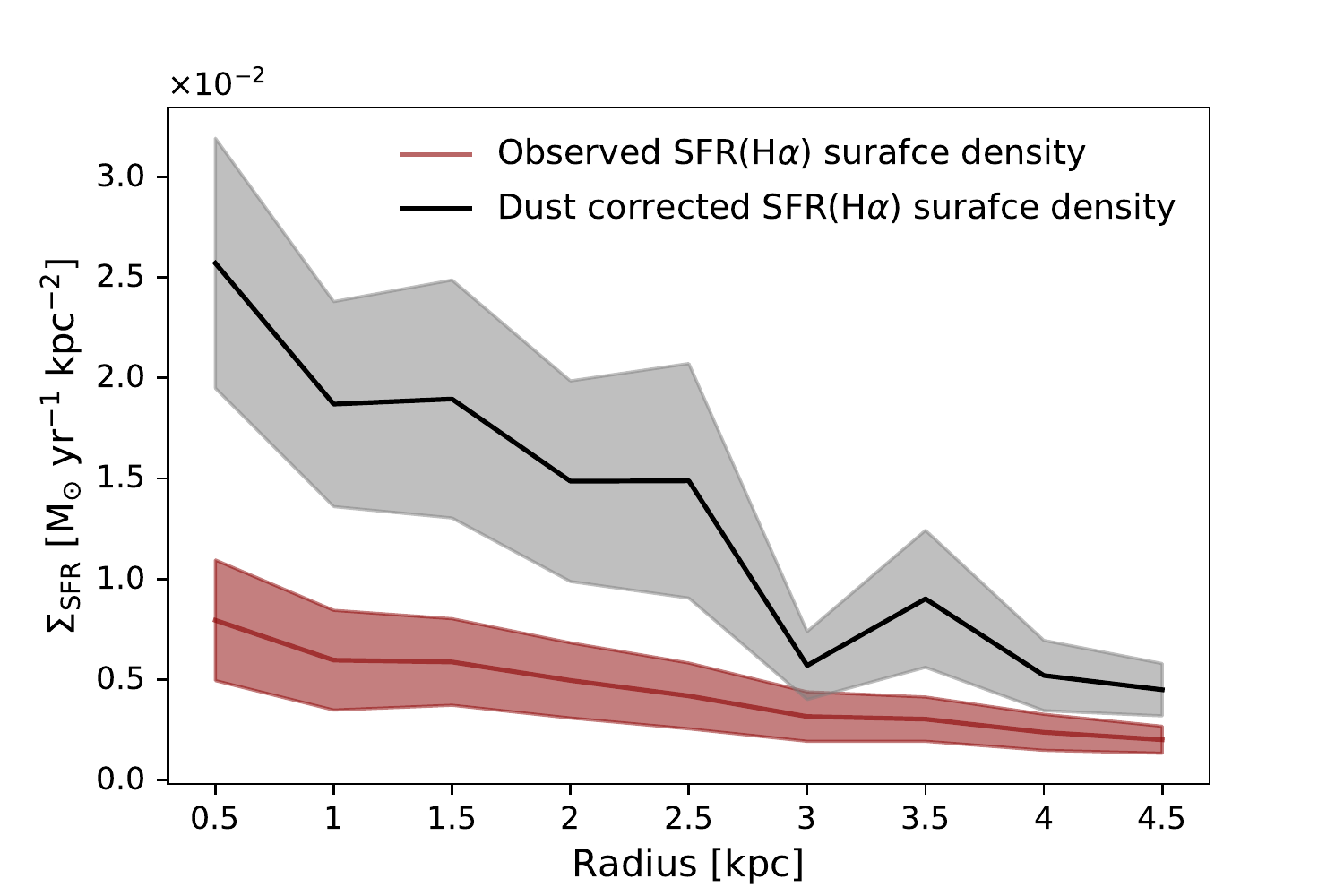}
\caption{Median radial profile of $\Sigma_{\mathrm{SFR}}$ for all the galaxies in the sample. Red and black profiles represent the SFR$_\mathrm{H\alpha}$, and the dust-corrected SFR$_\mathrm{H\alpha}$, respectively. Shaded regions correspond to 1$\sigma$ error in the data. }
\label{fig: SFR profile}
\end{figure}

The radial profile of the stellar mass surface density declines smoothly from the most inner regions outwards to the disk, consistent with previous studies (e.g., \citealt{Wuyts2012,Hemmati2015}) at similar redshifts. The radial profile of the SFR also declines towards outer parts of galaxies, but not as smooth as the mass density. In Figure \ref{fig: SFR profile} we show the radial profile of the median SFR surface density ($\Sigma_{\mathrm{SFR}}$) over all the galaxies in our sample out to 4.5 kpc radius. Shaded regions in this figure represent the standard deviation over the full sample. Red profile shows the observed SFR$_{\mathrm{H}\alpha}$ (uncorrected for dust), increasing towards the inner regions of galaxies with a relatively shallow slope. However, the increased SFR in the central regions become clearer when SFR$_{\mathrm{H}\alpha}$ is corrected for dust using the Balmer decrement (black profile). This also suggests that dust is more centrally concentrated, and decreases towards the outskirts, consistent with previous results (e.g., \citealp{Hemmati2015,Nelson2016(1)}).

\citet{Nelson2016(1)} used resolved measurements of H$\alpha$ and H$\beta$ emission lines for $\sim$600 galaxies at $\textit{z}\sim$ 1.4 to study the radial distribution of SFR$_{\mathrm{H}\alpha}$ and dust attenuation. They found high central SFRs and steeper dust attenuation gradients with increasing stellar mass of galaxies, such that H$\alpha$ attenuation (A$_{\mathrm{H}\alpha}$) in the center of galaxies with the average mass of log M$_{*}/\mathrm{M}_{\odot}=10.2$ is up to 2 magnitudes. However, for galaxies with log M$_{*}/\mathrm{M}_{\odot}=9.2$ they observed little dust attenuation at all radii, whereas according to Figure \ref{fig: SFR profile}, dust obscures the SFR by a factor of $\sim$ 3 at the very center of the galaxies in our sample with the average mass of M$_{*}= 10^{8.71}\mathrm{M}_{\odot}$. In the next subsection we look in more detail into the distribution of dust inside galaxies in our sample.

\begin{figure}
\includegraphics[width=0.47\textwidth]{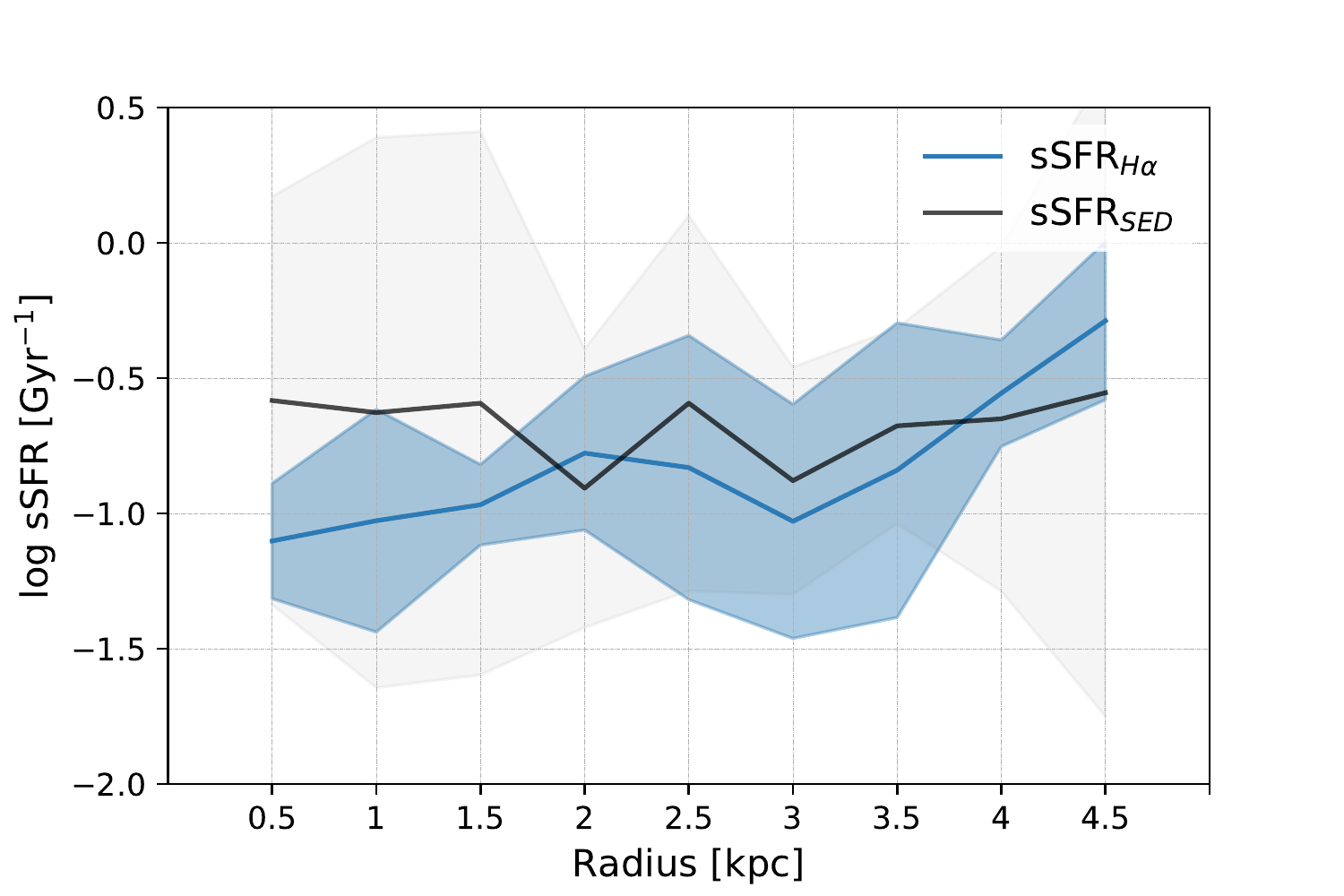}
\caption{Median radial sSFR profiles for galaxies in the sample. Solid black profile presents the SED-derived sSFR and the solid blue profile shows the sSFRs calculated by dividing the dust-corrected SFR$_{\mathrm{H}\alpha}$ by the SED-derived stellar mass in each annulus. In both profiles, shaded regions represent 1$\sigma$ error in the data. }
\label{fig:sSFR profile}
\end{figure}

To better investigate the mass build up in these galaxies, we also present the radial profile of the specific star formation rate (sSFR) in Figure \ref{fig:sSFR profile}. Here, resolved sSFRs are estimated both by pixel-by-pixel SED-fitting (grey line/ shaded region) and dividing the dust-corrected SFR$_{\mathrm{H}\alpha}$ by the SED-derived stellar mass (blue line/shaded region). In the second approach, the median mass and median SFR of the pixels inside a ring is used to estimate the sSFR per annulus. The shaded regions in this Figure represent the standard deviation of sSFRs at different radii in our sample. As expected, standard deviation is much larger for the sSFR$_{\mathrm{SED}}$ profile at all radii which is due to uncertainties associated with SFR$_{\mathrm{SED}}$. The radial profile of the sSFR$_{\mathrm{SED}}$ is almost flat throughout the galaxy; however, sSFR$_{\mathrm{H}\alpha}$ shows $\sim$ 0.8 dex increase in median sSFR from center out to 4.5 kpc radii.

Nearly flat trend for SED-based sSFR was previously observed by \citet{LiuSandy} for star forming galaxies with the mass ranging from $10^{9}$ to $10^{10}$ M$_{\odot}$, whereas they found an increasing sSFR profile for galaxies in their most massive bin (M$_{*}$ $>$ $10^{10.5}$ M$_{\odot}$). \citealt{BelfioreManGa2018} also found a mostly flat sSFR profile based on H$\alpha$ measurements from MaNGA survey for the lowest mass galaxies in a sample of star forming main sequence and green valley galaxies ($10^{9}$ to $10^{9.5}$ M$_{\odot}$), and an increasing sSFR profile for higher mass galaxies. However, Figure \ref{fig:sSFR profile} suggests that sSFR$_{\mathrm{H}\alpha}$ profile is not necessarily flat in our sample with the average mass of M$_{*}= 10^{8.71}\mathrm{M}_{\odot}$. This increase of sSFR with radius implies that the central regions of these galaxies formed at earlier times such that they are less gas rich compared to the outer regions, which suggests an inside-out growth scenario for the formation of these galaxies. 
We note that the sample size here is small and the scatter among the galaxies (visible from the shaded regions in Figure\ref{fig:sSFR profile}) is large to draw strong conclusions. However, the flatness of sSFR$_{\mathrm{SED}}$ compared to the decline of the sSFR$_\mathrm{H\alpha}$ towards the central parts of the galaxy, also suggests an inside-out quenching of star formation. This is again due to the different timescales of SFR that H$\alpha$ and SED are probing, where per unit stellar mass, there is less recent SFR compared to the longer timed average SFR in inner regions of galaxies compared to outer regions in the disk.

\vspace{3.75mm}

\begin{figure*}[!hbt]
\includegraphics[width=\textwidth]{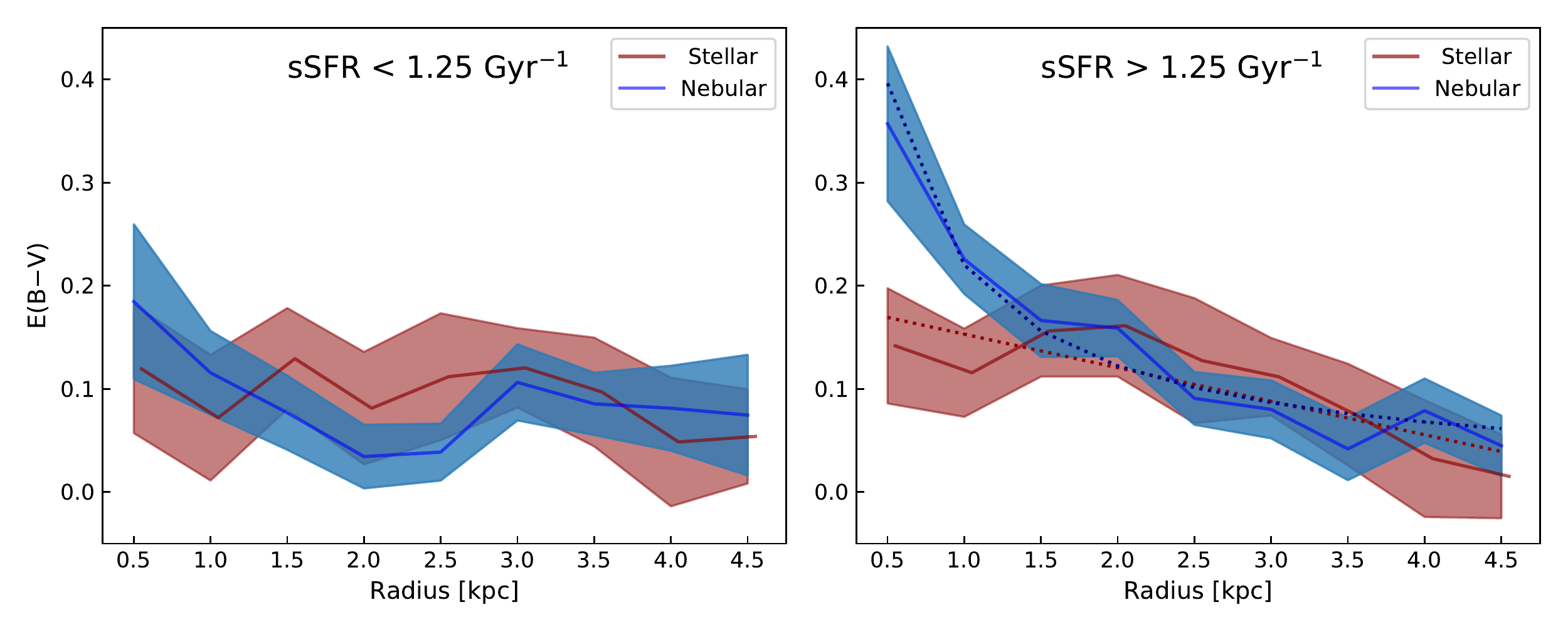}
\caption{Median radial profiles of nebular and stellar color excess for galaxies in the sample. We divided our sample into two bins based on their integrated sSFR such that the number of galaxies in two bins are about the same. Galaxies in the left panel have integrated sSFR$<1.25$ Gyr$^{-1}$, and galaxies in the right panel have integrated sSFR$>1.25$ Gyr$^{-1}$. In each panel, red and blue profiles present the stellar and nebular color excess, respectively. In all profiles, shaded regions represent 1$\sigma$ scatter in the data. In the right panel, the dotted lines correspond to the best fit model to the data where the stellar reddening is modeled with a linear relation and nebular profile is fitted by a power-law.}
\label{fig:radialebv}
\end{figure*}

\vspace{7mm}
\subsection{Extinction profiles}

In this subsection we present the radial profiles for the stellar and nebular extinctions. For this purpose, resolved stellar E(B$-$V) is estimated through pixel-by-pixel SED-fitting and nebular E(B$-$V) is measured from the H$\alpha$ and H$\beta$ emission lines in every pixel following Equation \ref{eq:ebv}:

\begin{equation}
E(B-V)_{\mathrm{gas}}=\frac{2.5}{k(H\beta)-k(H\alpha)}\log_{10}(\frac{H\alpha/H\beta}{2.86})
\label{eq:ebv}
\end{equation} 

where k(H$\alpha$)=3.326 and k(H$\beta$)=4.598, are values of extinction curve at the wavelength of these two lines computed based on Calzetti extivtion curve (\citealt{CALZETTI2000}), and 2.86 is the intrinsic Balmer decrement for Case B recombination at T = 10000 K and electron density of $10^4$ cm$^{-3}$. The median E(B$-$V) in each annulus is used to produce radial dust profiles in individual galaxies. We then extracted reddening profile of the sample by taking the median E(B$-$V) at each annulus over all the galaxies. 
The radial profiles of both nebular and stellar dust vary from galaxy to galaxy in our sample, so defining one dust profile shape for the whole sample is not valuable. However, integrated properties of galaxies are known to be correlated with dust inside galaxies (e.g., \citealt{Reddy2015}). We therefore, study the dependence of the reddening profiles on integrated physical properties including stellar mass, SFR and sSFR and found that the profiles are most significantly dependent on the integrated sSFRs.

Figure \ref{fig:radialebv} presents the nebular and stellar E(B$-$V) profiles when we divided our sample into two sSFR bins with similar number of galaxies in each bin. At the lower sSFR bin (left panel of Figure \ref{fig:radialebv}), the Pearson correlation coefficients between E(B$-$V) and radius is within the range of 0 and -0.5 showing a weak radial dependence of both stellar and nebular reddening which can be explained by low values of extinction at all radii for these galaxies. On the contrary, in the higher sSFR bin (right panel of Figure \ref{fig:radialebv}), the correlation coefficient of both nebular and stellar reddening and radius are less than -0.8, indicative of a strong relation between E(B$-$V) and radius. We fitted a power-law to the nebular profile as E(B$-$V)= A$r^b$ where A=$0.22\pm 0.02$ and b=$-0.85\pm 0.10$ (blue dotted line in Figure \ref{fig:radialebv}) , and a linear relation to the stellar reddening as E(B$-$V)= mr + c where m=$-0.03\pm 0.01$ and c=$0.19\pm 0.02$ (red dotted line in Figure \ref{fig:radialebv}).

Clear from Figure \ref{fig:radialebv} and as already seen in section \ref{sec:sfr profile}, the highest attenuation by dust occurs in the central regions of the galaxies with the exception of stellar E(B$-$V) in low integrated sSFR galaxies. These plots also indicate that sSFR is affecting both nebular and stellar E(B$-$V) profiles such that galaxies with higher integrated sSFR, have higher values of both types of reddening at all radii. Also, galaxies in the higher sSFR bin tend to have steeper profiles, showing higher nebular and stellar reddening toward the center of these galaxies. 

Another essential notion to examine in our sample is the extra attenuation towards nebular emission lines compared to the stellar light. The ratio of nebular to stellar reddening for the integrated light of galaxies has been extensively studied (e.g., \citealp{CALZETTI2000, Reddy2010, wild2011,Koyama2019}), there is however a large scatter in this relation. \citet{Reddy2015} related this scatter to physical properties such as sSFR and stellar mass based on observations of star-forming galaxies at \textit{z}$\sim$2. \citet{Hemmati2015} presented the nebular to stellar ratio at kpc-scales in a small sample of emission line galaxies at z $\sim$ 0.4, and also found extra reddening towards the nebular emission lines, increasing with the mass surface density. This is consistent with the overall trend seen in our sample, visible in Figure \ref{fig:radialebv} where the blue shaded region sits above the red one in inner regions of galaxies with higher mass surface densities.

While the trends of nebular and stellar dust reddening over the whole sample agrees with what has been suggested before, the nebular to stellar reddening ratio in individual galaxies seems unique for every single galaxy. This is expected given the sensitivity of the profiles to the patchiness of dust and the distribution of clumps in galaxies, so a unified trend is not observed for individual galaxies at kpc-scales. Also, the observed ratio of nebular to stellar reddening varies significantly among the galaxies. However, we find that the nebular to stellar color excess profile is closely following the sSFR radial profile such that for the younger parts of the galaxy with higher current SFR (high sSFR), nebular to stellar color excess is increasing significantly. Figure \ref{fig:ebvsingle} in the appendix, presents the radial profile of the nebular E(B$-$V) to stellar E(B$-$V) for three sample galaxies over plotted on their sSFR profiles. These are the same galaxies as the resolved kpc-scale maps in Figure \ref{fig:resolvedmaps} (galaxies from left to right in Figure \ref{fig:ebvsingle} correspond to those from top to bottom in Figure \ref{fig:resolvedmaps}). This figure allows visual inspection of correlations between the shape of reddening profiles and the resolved structures inside galaxies, such as the position of star forming clumps (e.g., galaxy on top panels of Figure \ref{fig:resolvedmaps}) or the non-central mass distribution (e.g., galaxy on bottom panels of Figure \ref{fig:resolvedmaps}).

\subsection{Resolved SFR-M$_{*}$ and sSFR-M$_{*}$ relations}

\begin{figure*}[!hbt]
\includegraphics[width=\textwidth]
{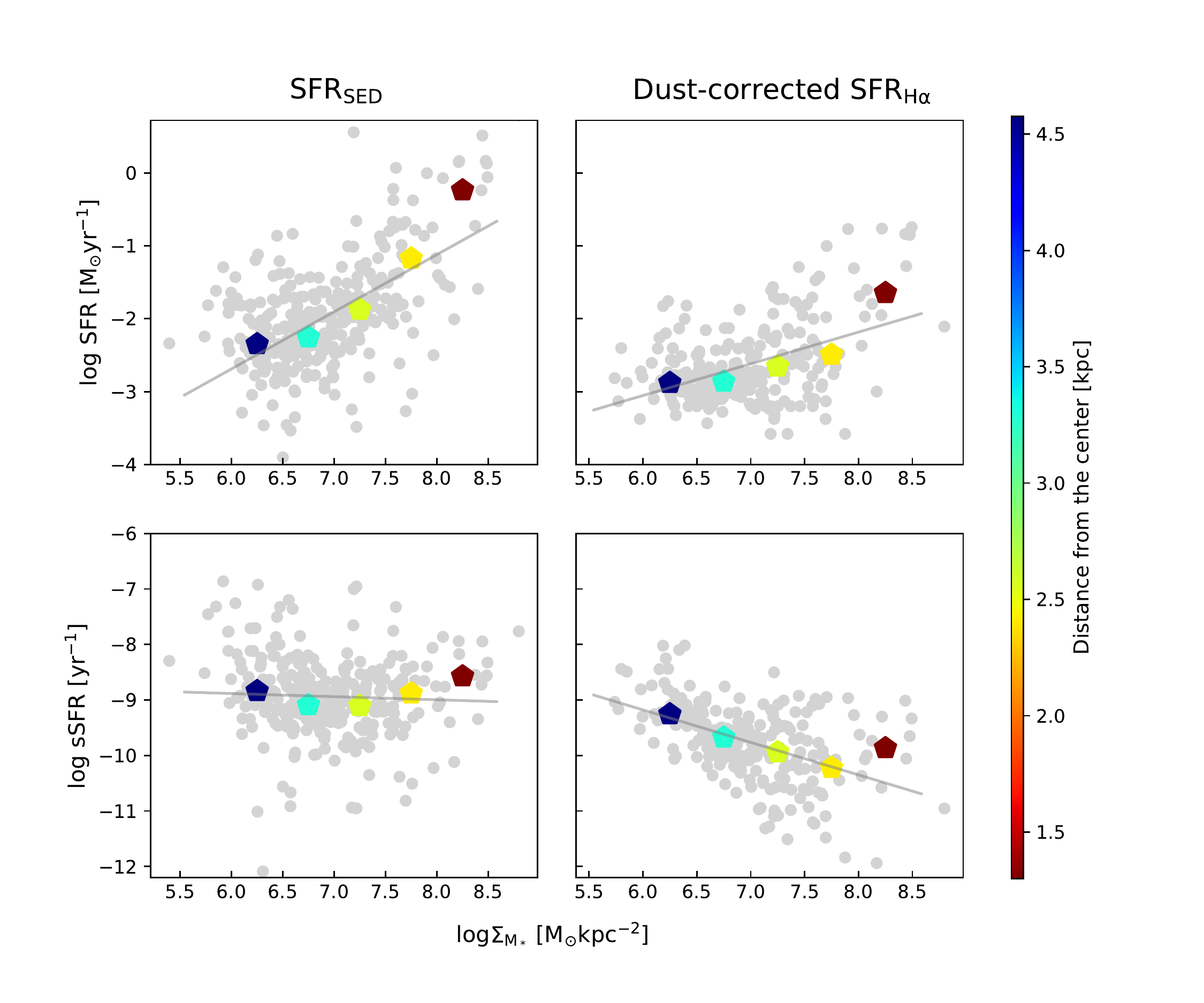}
\caption{Resolved main sequence and resolved sSFR-M$_{*}$ relations for galaxies in our sample. Gray data points represent individual annuli in galaxies. Two top panels are presenting resolved main sequence with two different SFR diagnostics. Left panel is based on SFR$_{\mathrm{SED}}$ and right panel is based on dust-corrected SFR$_{\mathrm{H\alpha}}$. Bottom panels present sSFR-M$_{*}$ relation with sSFR$_{\mathrm{SED}}$ on the left and dust-corrected sSFR$_\mathrm{H\alpha}$ on the right. Solid gray lines are the best linear fits to grey data points with the fitted parameters presented in Table \ref{table:2}. Colored pentagons are the median SFR/sSFR  (top/bottom panels) in bins of stellar mass which are color-coded by the average distance from the center of the galaxies.}
\label{fig:resolved main sequence}
\end{figure*}

 The main sequence of star forming galaxies is a well studied relation between integrated SFR and stellar mass of galaxies (e.g. \citealt{Noeske2007,Speagle2014,renzini2015}). This relation can also be studied at resolved scales to reveal how kpc-scale properties are related to unresolved measurements (e.g. \citealt{Wuyts2013,Hemmati2014,ellison2018}). Studying the sSFR-M$_{*}$  relation also reveals information about the star-formation history of galaxies. At small scales, this relation can be used to infer when different stellar populations were formed and to differentiate between galaxy growth scenarios using the radial gradient in sSFR-M$_{*}$ relation.

In Figure \ref{fig:resolved main sequence} we present the resolved main sequence, where the top panels show the resolved SFR versus mass surface density and bottom panels showing resolved sSFR versus mass surface density for all the galaxies in our sample. On the left panels, SFR is calculated from SED fitting and right panels are based on dust-corrected SFR$_{\mathrm{H\alpha}}$. Each grey data point on these plots represents an annulus of a galaxy with the width of 1 kpc, where the median value of the $\Sigma_{\mathrm{M}_{*}}$, $\Sigma_{\mathrm{SFR}}$ and sSFR of the pixels are assigned to that annulus. Colored pentagons in these plots show the median SFR/sSFR in mass bins with the width of 0.5 dex, color coded by the distance from the center. We perform least square regression to grey data points shown as solid gray lines with the best fitted parameters presented in table \ref{table:2}. In all panels, the best fit line is a good representation of the median SFR/sSFR in each mass bin (colored pentagons) except for the highest mass bin which is most likely due to poor statistics. 

The top two panels confirm that linear relation between SFR and M$_{*}$ (so-called main sequence) exists at resolved scales with different slopes and normalization depending on the SFR diagnostics. This slope is much steeper when SFRs are derived from SED-fitting (0.78 in compare to 0.43) suggesting that massive regions of galaxies have already built up their stellar mass through forming stars over their lifetime. Recent star formation traced by SFR$_{\mathrm{H\alpha}}$ is relatively lower in these massive regions. We also infer that the position of different parts of galaxies on this plot is strongly dependent on their radial distance such that inner regions (redder points) have higher $\Sigma_{\mathrm{M}_{*}}$ and $\Sigma_{\mathrm{SFR}}$. 

\begin{table}[!hbt]
\caption{\label{table:2} Best fit values to the resolved SFR-M$_{*}$ and sSFR-M$_{*}$ relations}
\begin{tabular}{cccc} 
\hline
Relation & Slope & y-intercept & RMSE \\
\hline
\\
SFR$_{\mathrm{SED}}$-M$_{*}$ & 0.78 & -7.39 & 0.09\\
SFR$_{\mathrm{H\alpha}}$-M$_{*}$ & 0.43 & -5.66 & 0.05\\
\\
\hline
\\
sSFR$_{\mathrm{SED}}$-M$_{*}$ & -0.05 & -8.55 & 0.09\\
sSFR$_{\mathrm{H\alpha}}$-M$_{*}$& -0.59& -5.67 & 0.07\\
\\
\hline
\end{tabular}
\centering
\end{table}

The bottom panels indicate that the sSFR$_{\mathrm{SED}}$ results in a shallower trend compared to sSFR$_{\mathrm{H\alpha}}$ (-0.05 vs. -0.59) which means that the radial gradient in recent SFR traced by H$\alpha$ is higher in comparison with average SFR traced by SED-fitting. In other words, massive regions which are already quenched show less evidence for recent star formation activity. Also, inner parts of the galaxies (redder points) have higher $\Sigma_{\mathrm{M}_{*}}$ and lower sSFR, implying that these parts contain the bulge of the galaxy and formed their stars before the outer regions (i.e. disks), which is again an evidence for inside-out growth of these galaxies.

\section{Summary}

In this work, we investigate the resolved photometric and spectroscopic properties of 32 galaxies at $0.1< z< 0.42$ with stellar masses ranging from $10^{7.7}$ to $10^{10.3}$ M$_\odot$. Following are the main points of this study:
\begin{itemize}

    \item We observe a linear relationship between SFR$_{\mathrm{SED}}$ and dust-corrected SFR$_{\mathrm{H}\alpha}$ with a near unity slope at lower SFRs, and deviation from unity at higher SFR values which can be explained by the effect of dust and/or quenching mechanisms which start in the more massive galaxies of our sample.
    
    \item We measure an increase of $\sim$ 0.8 dex in the median SFR$_{\mathrm{H\alpha}}$ from center to 4.5 kpc radii for galaxies in our sample. This suggests that sSFR is not necessarily constant with radius in relatively low-mass galaxies (M$_{*}< 10^{10}$ M$_{\odot}$) and inside-out growing is a possible scenario for the evolution of these galaxies.
    
    \item We find that radial profile of stellar and nebular E(B$-$V) are strongly dependent on the integrated sSFR of the galaxy. Radial profile of E(B$-$V) in galaxies with lower sSFR, are weakly dependent on radius; however, in galaxies with high sSFR, both stellar and nebular reddening have significantly higher values of extinction in inner parts of galaxies. We also study the radial profile of the ratio of nebular E(B$-$V) to stellar E(B$-$V) which found to be patchy and unique for individual galaxies with no specific trend with radius in this sample.
    
    \item We show that the slope and normalization of the resolved SFR-M$_{*}$ and sSFR-M$_{*}$ relations are highly dependent on the SFR diagnostics (SFR$_{\mathrm{SED}}$ vs. SFR$_{\mathrm{H\alpha}}$). Also,  the position of different regions of galaxies on these two plots is determined by their radial distance from the center in a way that central regions have higher $\Sigma_{\mathrm{M}_{*}}$ and $\Sigma_{\mathrm{SFR}}$ and lower sSFR, which implies that these parts formed their stars before the outer parts and is an another evidence for inside-out growth of these galaxies.

\end{itemize}
 
 We developed the methodology of studying resolved properties of galaxy samples with combined \textit{HST}+MUSE data. In future work we will incorporate the upcoming data from the completed MUSE-Wide Survey, which include $\sim 4$ times the sample discussed here to build upon our current analysis (data for 44 out of 100 fields of this survey is now available at \url{https://musewide.aip.de/project/}). Clearly, next generation telescopes and instruments will also allow us to study the spatially resolved properties of large and diverse samples of intermediate and high redshift galaxies with unprecedented detail.

\section*{Acknowledgement}
Part of this work is based on observations taken by the MUSE-Wide Survey as part of the MUSE Consortium (\citealt{MUSE2}). We use the MUSE Python Data Analysis Framework in some parts of our analysis as well (MPDAF, \citealt{mpdaf1,mpdaf2}). We also want to thank the anonymous referee for constructive suggestions and comments.

\bibliography{resolved_muse.bib}

\appendix
In this section, we bring spatially resolved mass, SFR, stellar E(B$-$V), H$\alpha$ and H$\beta$ maps for three sample galaxies, following by their radial profile of nebular to stellar E(B$-$V) and sSFR. These maps are examples of how distribution of physical properties can be so diverse and non-Gaussian, and allow visual inspection of how resolved structures of galaxies affecting the radial profile of sSFR and nebular to stellar reddening.
\vspace{4mm}

\begin{figure*}[h]
\centering
\includegraphics[scale=0.48]{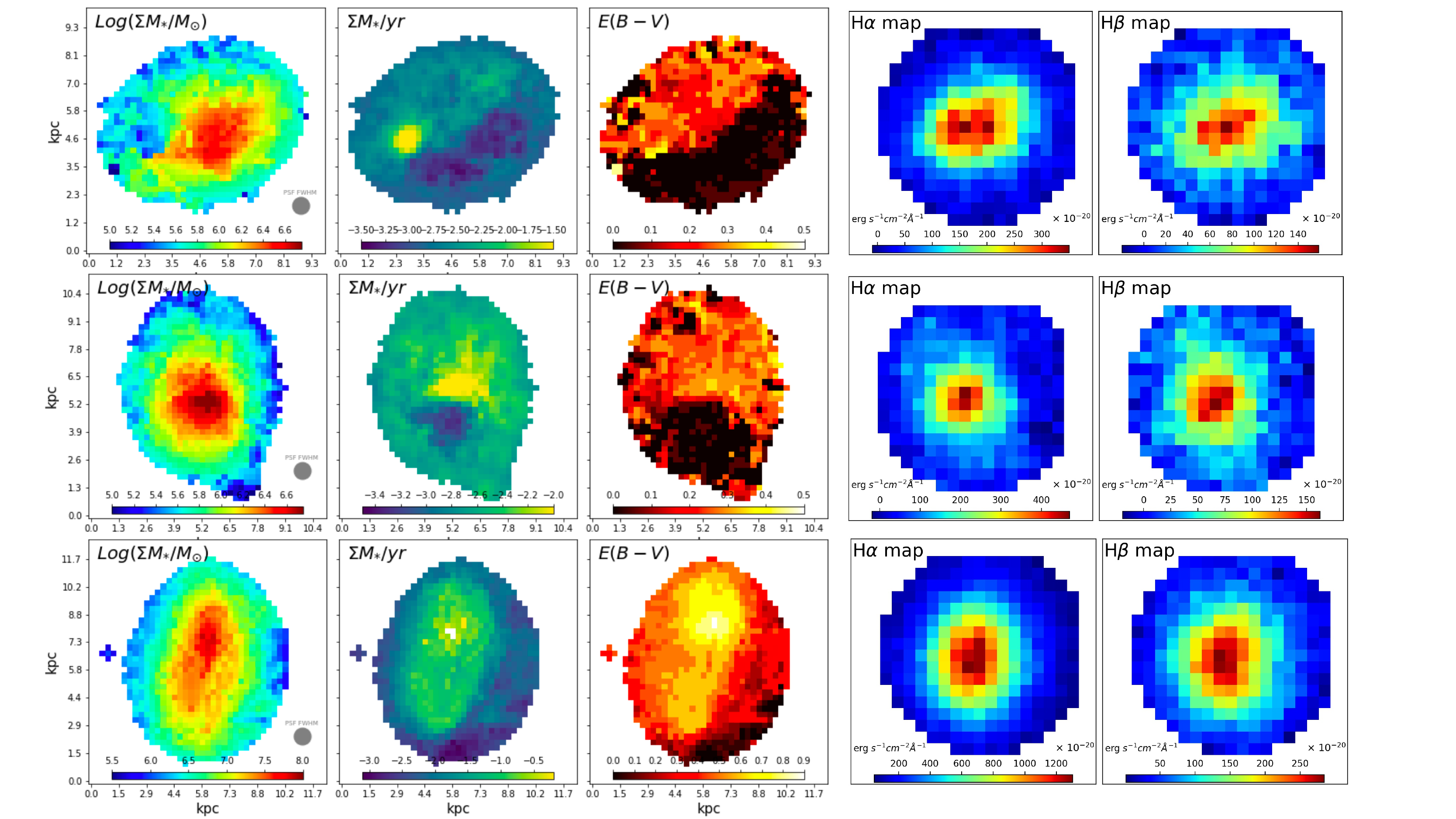}
\caption{From left to right: spatially resolved stellar mass surface density, SFR surface density, stellar E(B$-$V), H$\alpha$ emission and H$\beta$ emission maps for three galaxies from our sample. To allow for visually comparing \textit{HST} and MUSE PSF, resolved SED-fitting maps (first three in each row) have \textit{HST} quality where input seven images of SED-fitting were PSF matched to the resolution in F160W, and H$\alpha$ and H$\beta$ maps which are derived from 3D datacubes provided by MUSE-Wide Survey have their original PSF.}
\label{fig:resolvedmaps}
\end{figure*}

\begin{figure*}[h]
\centering
\includegraphics[scale=0.35]{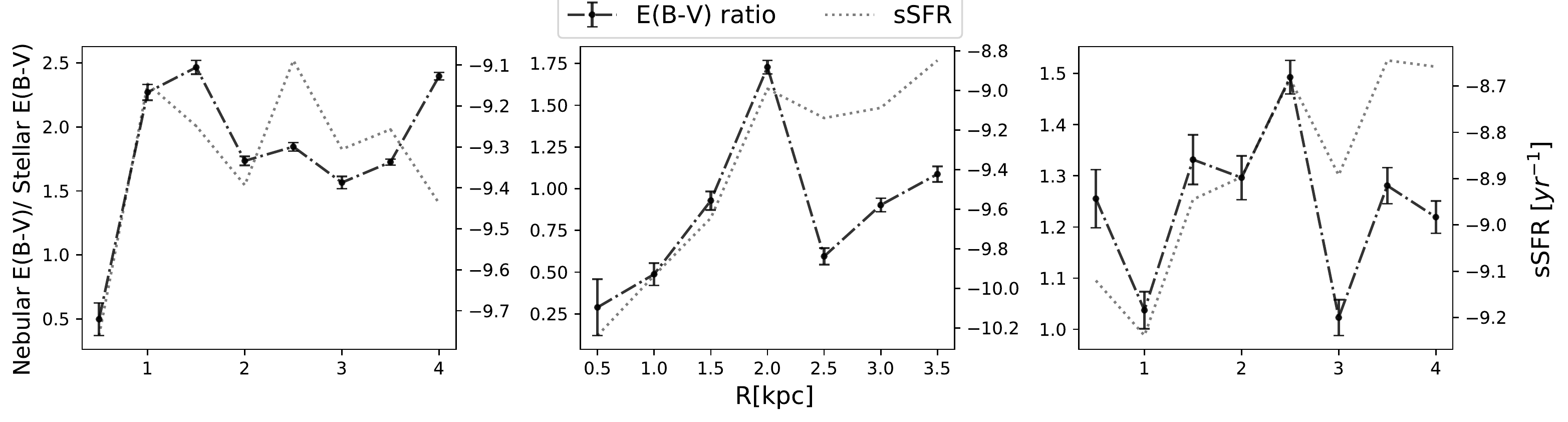}
\caption{Radial profile of nebular to stellar color excess (black dashed lines) overplotted on sSFR profile (grey dotted lines) for 3 sample galaxies which their nebular E(B$-$V) to stellar E(B$-$V) ratio profiles closely follow their sSFR profiles. The mass, SFR, stellar E(B$-$V), H$\alpha$ and H$\beta$ maps for these three galaxies are presented in Figure \ref{fig:resolvedmaps} (the top panel of figure \ref{fig:resolvedmaps} is corresponding to left panel of this graph).}
\label{fig:ebvsingle}
\end{figure*}

\end{document}